\documentclass[english]{article}
\usepackage[T1]{fontenc}
\usepackage[latin9]{inputenc}
\usepackage{geometry}
\usepackage{float}
\usepackage{amstext}
\usepackage{amssymb}
\usepackage{graphicx}
\usepackage{setspace}
\usepackage{esint}
\usepackage[authoryear]{natbib}
\makeatletter

\providecommand{\tabularnewline}{\\}
\floatstyle{ruled}
\newfloat{algorithm}{tbp}{loa}
\providecommand{\algorithmname}{Algorithm}
\floatname{algorithm}{\protect\algorithmname}

\usepackage{ntabbing}

\makeatother

\usepackage{babel}
\begin{document}

\title{Adaptive step size selection for Hessian-based manifold Langevin
samplers%
\thanks{The author is indebted to the reviewers, Hans A. Karlsen, Hans J.
Skaug, Anders Tranberg and participants at the 18th meeting of the
Norwegian Statistical Association for valuable comments.%
}}

\author{Tore Selland Kleppe%
\thanks{University of Stavanger, tore.kleppe@uis.no%
}}
\maketitle
\begin{abstract}
The usage of positive definite metric tensors derived from second
derivative information in the context of the simplified manifold Metropolis
adjusted Langevin algorithm (MALA) is explored. A new adaptive step
size procedure that resolves the shortcomings of such metric tensors
in regions where the log-target has near zero curvature in some direction
is proposed. The adaptive step size selection also appears to alleviate
the need for different tuning parameters in transient and stationary
regimes that is typical of MALA. The combination of metric tensors
derived from second derivative information and adaptive step size
selection constitute a large step towards developing reliable manifold
MCMC methods that can be implemented automatically for models with
unknown or intractable Fisher information, and even for target distributions
that do not admit factorization into prior and likelihood. Through
examples of low to moderate dimension, it is shown that proposed methodology
performs very well relative to alternative MCMC methods.
\end{abstract}

\section{Introduction}

Suppose $\tilde{\pi}(\mathbf{x})$ is a target density kernel where
$\mathbf{x}\in\mathbb{R}^{d}$ so that $\pi(\mathbf{x})=\tilde{\pi}(\mathbf{x})/\int\tilde{\pi}(\mathbf{x})d\mathbf{x}$
is a probability density function. Bayesian analysis of many statistical
models necessitates the calculation of integrals with respect to $\pi(\mathbf{x})$
when $\pi(\mathbf{x})$ is analytically intractable and only the unnormalized
density kernel $\tilde{\pi}(\mathbf{x})$ is available for evaluation
\citep[see e.g.][]{GelmanBDA3}. To approximate such integrals, Markov
Chain Monte Carlo (MCMC) has seen widespread use, and the development
of better MCMC algorithms that can help researchers tackle even larger
and more challenging models is still a highly active field \citep[see e.g.][]{RSSB:RSSB736,mcmc_handbook,girolami_calderhead_11}. 

MCMC relies on constructing a discrete time Markov process $\left\{ \mathbf{x}_{t}\right\} $
that has $\pi(\mathbf{x})$ as its stationary distribution. Consequently,
an ergodic theorem ensures that, under some technical conditions,
integrals on the form $\int g(\mathbf{x})\pi(\mathbf{x})d\mathbf{x}$
can approximated as $\hat{\mu}_{T}(g)=\frac{1}{T}\sum_{t=1}^{T}g(\mathbf{x}_{t})$
for some large $T$ \citep[e.g.][chapter 6]{robert_casella}. A very
general way of constructing such a Markov process, and that serves
as the foundation of most MCMC methods, is the Metropolis Hastings
(MH) algorithm \citep{Metropolis53,Hastings70}. Suppose the current
state of the Markov process is $\mathbf{x}_{t}$ and that $q(\cdot|\mathbf{x}_{t})$
is some proposal probability density function. Then the MH algorithm
proceeds by proposing $\mathbf{x}^{*}\sim q(\cdot|\mathbf{x}_{t})$,
and accepting the new state $\mathbf{x}_{t+1}$ to be equal to $\mathbf{x}^{*}$
with probability $\alpha(\mathbf{x}_{t},\mathbf{x}^{*})=\min\left[1,\tilde{\pi}(\mathbf{x}^{*})q(\mathbf{x}_{t}|\mathbf{x}^{*})/(\tilde{\pi}(\mathbf{x}_{t})q(\mathbf{x}^{*}|\mathbf{x}_{t}))\right]$.
With remaining probability $1-\alpha(\mathbf{x}_{t},\mathbf{x}^{*})$,
$\mathbf{x}_{t+1}$ is set equal to $\mathbf{x}_{t}$. The MH algorithm
is extremely general as only minimal restrictions on the proposal
density are necessary for $\left\{ \mathbf{x}_{t}\right\} $ to satisfy
detailed balance with $\pi(\mathbf{x})$ as the stationary distribution
\citep{robert_casella}.

These minimal restrictions have lead to a large number of strategies
for choosing proposal distributions \citep[see e.g.][]{liu_mc_2001,robert_casella}.
The most widespread strategy is the Gaussian random walk proposal,
corresponding to $q(\cdot|\mathbf{x}_{t})=\mathcal{N}(\cdot|\mathbf{x}_{t},\Sigma)$
where $\mathcal{N}(\mathbf{x}|\mu,\Sigma)$ denotes the $N(\mu,\Sigma)$
density evaluated at $\mathbf{x}$. The covariance matrix $\Sigma$
should be chosen according to the target distribution at hand. As
a rule of thumb, a scale of $q(\cdot|\mathbf{x}_{t})$ that is ``large''
relative to the scale of the target density leads to a low acceptance
probability, and thereby a slow exploration of target distribution.
At the other end of the spectrum, if the scale of $q(\cdot|\mathbf{x}_{t})$
is ``small'' relative to the scale of the target, one typically
obtain a high acceptance probability but the cumulative distance traversed
by many accepted steps will still be small. Provided that the variance
of $\hat{\mu}_{T}(g)$ exists, both extreme cases lead to the variance
of $\hat{\mu}_{T}(g)$ being large for fixed $T$, whereas intermediate
choices of scale typically lead to more efficient sampling in the
sense that the variance of $\hat{\mu}_{T}(g)$ is smaller for fixed
$T$ than the variance at the extreme cases. It should be noted that
``large'' and ``small'' scales relative to the target are not
absolute or even well-defined magnitudes, and depend among many other
aspects, on the dimension $d$ of the target \citep[see e.g.][]{Gelman95,roberts2001}.
An additional complication is that many target distributions of interest,
e.g. joint posterior distributions of latent variables and variance
parameter of the latent variables, contain significantly different
scaling properties in different regions of the support of the target.
Consequently, there is often a need for proposal distributions to
adapt to local properties of the target distribution to achieve efficient
sampling.

Many researchers have explored MCMC methods that exploit local derivative
information from the target distribution to obtain more efficient
sampling. Early contributions in this strand of the literature includes
the Metropolis adjusted Langevin algorithm (MALA) \citep[see e.g.][]{roberts1996,roberts_stramer_02}.
MALA is a special case of the MH algorithm where the proposal distribution
is equal to a time-discretization of a Langevin diffusion, where the
Langevin diffusion has the target distribution as its stationary distribution.
The drift term in such diffusion, and consequently the mean of the
proposal distribution, will depend on the gradient of $\log\tilde{\pi}(\mathbf{x})$.
To achieve reasonable acceptance rates, it is often necessary to choose
the (time-) step size of the proposal distribution rather small, and
it has been argued that the MALA in this case will revert to a behavior
similar to that of the random walk MH \citep{1206.1901}.

Another way of constructing MCMC samplers that exploit derivative
information is Hamiltonian (or Hybrid) Monte Carlo (HMC) \citep{Duane1987216,liu_mc_2001,1206.1901,beskos2013,1411.6669,JMLR:v15:hoffman14a}.
HMC has the potential of producing proposals that are far from the
current state, while retaining arbitrarily high acceptance rates.
The proposals originate from numerically integrating a set of ordinary
differential equations that involve the gradient of $\log\tilde{\pi}(\mathbf{x})$
and it is often necessary to evaluate this gradient hundreds of times
per proposed state to achieve reasonable acceptance rates while retaining
proposals that are far from the current state.

Most implementations of MALA and HMC involve a user-specified scaling
matrix (in addition to other tuning parameters) to achieve reasonable
efficiency. In a landmark paper, \citet{girolami_calderhead_11} proposes
to recast MALA and HMC on a Riemann manifold that respects the local
scaling properties of the target, and consequently alleviates the
need to choose a global scaling matrix. Suppose $\mathbf{x}$ are
the parameters to be sampled and $\mathbf{y}$ the observations associated
with a Bayesian statistical model that admit the explicit factorization
into data likelihood $p(\mathbf{y}|\mathbf{x})$ and prior distribution
$p(\mathbf{x})$, i.e. $\tilde{\pi}(\mathbf{x})\propto p(\mathbf{y}|\mathbf{x})p(\mathbf{x})$.
Then \citet{girolami_calderhead_11} took their metric tensor to be
the matrix 
\begin{equation}
G_{GC}(\mathbf{x})=-E_{p(\mathbf{y}|\mathbf{x})}\left[\nabla_{\mathbf{x}}^{2}\log p(\mathbf{y}|\mathbf{x})\right]-\nabla_{\mathbf{x}}^{2}\log p(\mathbf{x}),\label{eq:GC-tensor}
\end{equation}
namely the Fisher information matrix associated with data likelihood
plus the negative Hessian of the log-prior. This choice was demonstrated
to be highly suited for Langevin and Hamiltonian based algorithms
that take local scaling properties into account. However, due to the
expectation in (\ref{eq:GC-tensor}), the Fisher information matrix
is rarely available in closed form, and the potential for automating
the implementation of MCMC samplers based on such a metric seems a
formidable task. 

In this paper, an alternative approach based on deriving a metric
tensor for simplified manifold MALA directly from the negative Hessian
of $\log\tilde{\pi}(\mathbf{x})$ is explored. As noted by \citet{girolami_calderhead_11}
and several of the discussants of that paper \citep[see e.g.][]{disc_sanzserna,disc_NTNU,disc_Jasra},
this approach admit a high degree of automation via e.g. the usage
of automatic differentiation software \citep{grie:2000}, but is complicated
by the fact that the negative Hessian is typically not positive definite
over the complete support of the target. Obtaining positive definite
local scaling matrices from potentially indefinite Hessian matrices
for application in modified Newton methods has seen substantial treatment
in non-convex numerical optimization literature \citep[see section 6.3 of][ for an overview]{noce:wrig:1999}.
Such approaches typically rely on either modification of the eigenvalues
via a full spectral decomposition of the Hessian, or some form of
modified Cholesky algorithm that produces a factorization of a symmetric
positive definite matrix that is close to the Hessian in some metric.
In the MCMC context, the former approach was taken by \citet{martin_etal2012,1212.4693},
whereas in the present paper I advocate the latter as it has potential
to exploit sparsity of the Hessian. 

The added generality with respect to target densities that can be
handled by Hessian-based metric tensors (based either on the spectral
decomposition or modified Cholesky factorizations) relative to (\ref{eq:GC-tensor})
comes at the cost of potential pathological behavior in regions of
the log-target with near zero curvature in some direction. To counteract
such effects, the primary contribution of this paper is a new adaptive
step size procedure for simplified manifold MALA that chooses the
step size locally based how well the Hessian-based metric tensor reflects
the local scaling of the target. The adaptive step size procedure
remains active throughout the MCMC simulation and at the same time
retains the target as the stationary distribution. The adaptive step
size procedure exploits the well known fact that a MALA update is
equivalent to a particular HMC method using one time integration step
\citep{1206.1901}, and therefore admits selection of step sizes for
simplified manifold MALA based on the (dimensionless) energy error
of the associated Hamiltonian system. An additional pro of the adaptive
step size procedure is that it appears to alleviate the need for different
tuning parameters in transient and stationary regimes that is typical
for standard MALA implementations \citep{RSSB:RSSB500}.

The remaining of the paper is laid out as follows. Section 2 describes
the applied metric tensor and the adaptive step size selection, and
illustrates the proposed methodology using a pilot example. Section
3 illustrates the proposed methodology on two realistic example models,
and compares the proposed methodology to alternative MCMC methods.
Finally section 4 provides some discussion.

\section{Adaptive step size modified Hessian MALA}

Assume that $\tilde{\pi}(\mathbf{x})$ is sufficiently smooth to admit
continuous and bounded derivatives up to order 2. I denote the gradient
of the log-kernel $g(\mathbf{x})=\nabla_{\mathbf{x}}\log\tilde{\pi}(\mathbf{x})$
and the Hessian of the log-kernel as $H(\mathbf{x})=\nabla_{\mathbf{x}}^{2}\log\tilde{\pi}(\mathbf{x})$.
The $d\times d$ identity matrix is denoted $I_{d}$. 

I take as vantage point the simplified manifold MALA (sMMALA) or position
specific preconditioned MALA of \citet{girolami_calderhead_11} \citep[see also][]{e16063074,Xifara201414},
which is a MH method characterized by the proposal distribution 
\begin{equation}
q(\cdot|\mathbf{x})=\mathcal{N}\left(\cdot|\mathbf{x}+\frac{\varepsilon^{2}}{2}\left[G(\mathbf{x})\right]^{-1}g(\mathbf{x})\;,\;\varepsilon^{2}\left[G(\mathbf{x})\right]^{-1}\right).\label{eq:sMMALA_prop_distr}
\end{equation}
Here $G(\mathbf{x})$ is a $d\times d$ symmetric and positive definite
matrix, from now on referred to as the metric tensor, and $\varepsilon$
is a tunable step size.

\subsection{Modified Cholesky based metric tensor\label{sub:The-GMW-factorization}}

This section explains a metric tensor based on an adaptation of the
modified Cholesky decomposition of \citet{gill_murray_1974}, \citet{gilletal},
from now on referred to as GMW. Suppose $A\in\mathbb{R}^{d\times d}$
is a symmetric, not necessarily positive definite matrix. Given a
user-specified small positive scale parameter $u$, our adaptation
of GMW (details and algorithm in Appendix \ref{sec:A-version-of-chol})
aims at finding a Cholesky decomposition $LL^{T}$ where $L$ is lower
triangular so that
\begin{equation}
\hat{A}=LL^{T}=A+J,\text{ where }L_{i,i}^{2}\geq u\max\left(1,\max_{1\leq i,j\leq d}|A_{i,j}|\right)>0,\; i=1,\dots,d.\label{eq:Def_LLchol}
\end{equation}
The matrix $J$ is a diagonal matrix with non-negative elements chosen
to ensure that $\hat{A}$ is positive definite, and in particular
the design of the GMW Cholesky algorithm ensures that if $A$ is sufficiently
positive definite, then $J=0$. As $J$ is a diagonal matrix and assuming
that $A$ has non-zero diagonal elements, the sparsity structures
of $A$ and $A+J$ are identical. This implies that whenever $A$
has a sparsity structure that can be exploited using sparse numerical
Cholesky factorizations \citep{davis_sparse}, the same applies for
the GMW factorization. 

The GMW factorization has seen widespread use within the numerical
optimization literature as a key ingredient in practical Newton optimization
algorithms for non-convex optimization problems \citep{noce:wrig:1999},
and has also served as the basis for further refinements as in \citet{doi:10.1137/0911064,doi:10.1137/S105262349833266X}.
For simplicity, I employ a variant of the algorithm given in \citet{gilletal}. 

In the remainder of this paper I work with the modified Hessian metric
tensor 
\[
G_{MC}(\mathbf{x})=L(\mathbf{x})L(\mathbf{x})^{T},
\]
where $L(\mathbf{x})$ is the output from the GMW Cholesky factorization
(\ref{eq:Def_LLchol}) applied to $-H(\mathbf{x})$. In itself, this
approach for finding a metric tensor is not fail-safe. To see this,
consider for instance the $d=1$ case, for which the proposed metric
tensor reduces to 
\begin{equation}
G_{MC}(\mathbf{x})=\max(u,|H(\mathbf{x})|).\label{eq:1D-metric}
\end{equation}
\begin{figure}
\centering{}\includegraphics[scale=0.8]{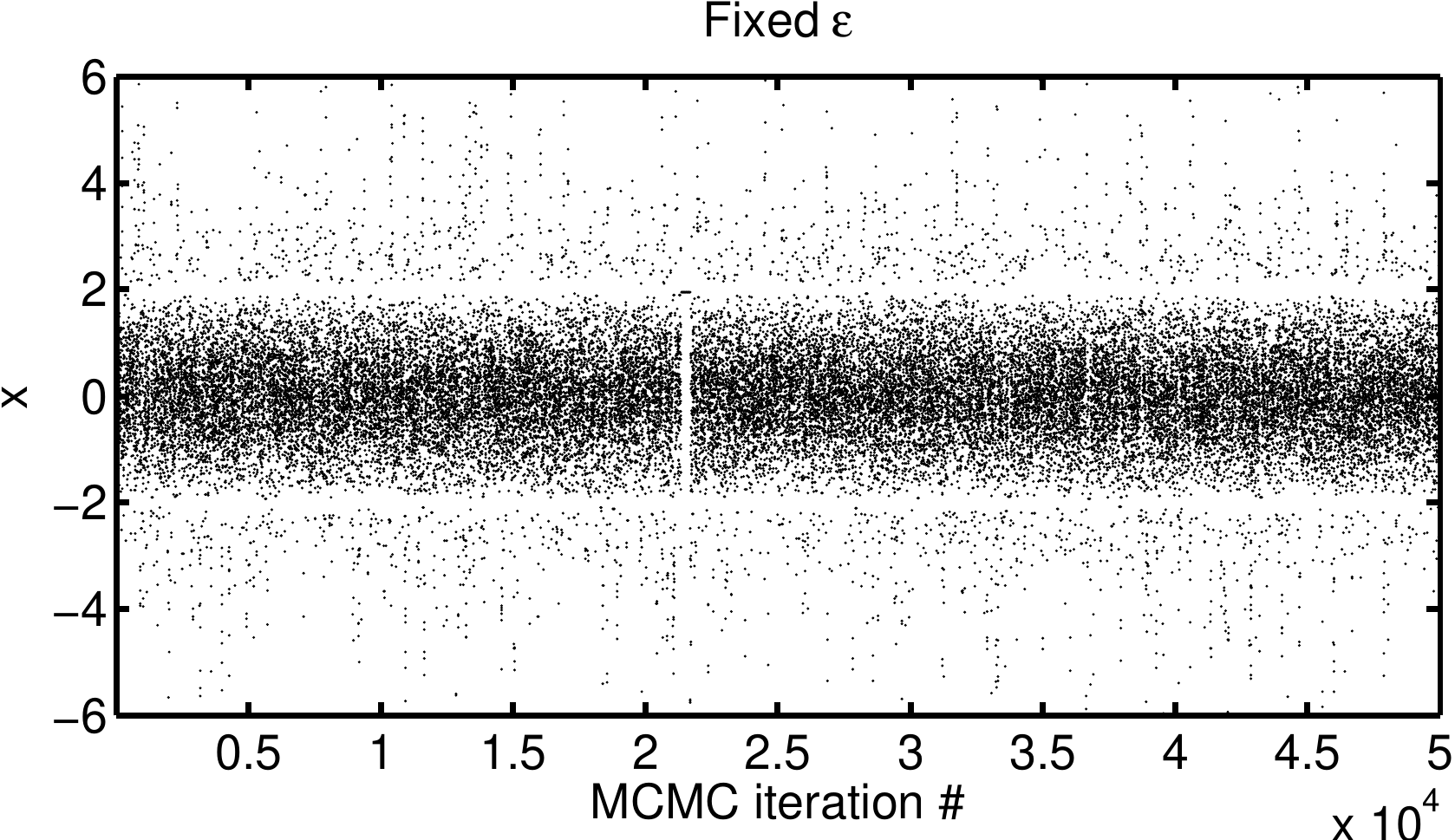}\protect\caption{\label{fig:fixed-trace-t}50,000 iterations of sMMALA with metric
tensor (\ref{eq:1D-metric}), $u=0.001$ and $\varepsilon=0.75$ applied
to a $t$-distribution with 4 degrees of freedom. It is seen that
the chain rarely enters the regions near $|\mathbf{x}|$=2, and when
it does, it tends to get stuck. An example of this is seen starting
in iteration 21347 where the chain is stuck for 340 iterations at
approximately 1.9529.}
\end{figure}
Typically I take $u$ to be rather small in order not to disturb the
proposal mechanism of sMMALA (\ref{eq:sMMALA_prop_distr}) in regions
where $|H(\mathbf{x})|$ is small but still contains useful information
on the local scaling properties of the target distribution. However
if left unmodified, a sMMALA method based on (\ref{eq:1D-metric})
would very infrequently move into regions near roots of the second
derivative (inflection points). To see this, observe that the backward
transition density $q(\mathbf{x}_{t}|\mathbf{x}^{\text{*}})$ occurring
in the nominator of the acceptance probability will be 
\begin{equation}
q(\mathbf{x}_{t}|\mathbf{x}^{\text{*}})=O(\sqrt{u}/\varepsilon)\text{ when }G(\mathbf{x}^{\text{*}})=u.\label{eq:alpha_order}
\end{equation}
When the method has moved into such a region, the chain will tend
to be stuck for many iterations as the proposal distribution then
has too large variance. 

To illustrate this phenomenon, consider a $t$-distributed target
with 4 degrees of freedom. In this case there are two inflection points
of the log-density located at $|\mathbf{x}|=2$. Figure \ref{fig:fixed-trace-t}
shows 50,000 MCMC iterations using sMMALA based on (\ref{eq:1D-metric})
for this target. It is seen that the chain rarely enters the regions
close to the inflection points, and when it does, the chain tend to
get stuck. It is worth noticing that these effects are not unique
to the modified Cholesky approach. Specifically, (\ref{eq:1D-metric})
would also be the result if the metric tensor was computed by modifying
the eigenvalues of a full spectral decomposition as in \citet{martin_etal2012},
and the metric tensor would be a soft/regularized absolute value applied
to $H(\mathbf{x})$ in the methodology of \citet{1212.4693}. Rather
the effects are the cost of the added generality associated with using
second order derivative information directly rather than via the Fisher
information matrix as in \citet{girolami_calderhead_11}. To mitigate
these undesirable effects, (\ref{eq:alpha_order}) suggest either
that $u$ or $\varepsilon$ must be chosen adaptively. In this work,
I focus on the latter as a simple and computationally efficient method
can be implemented for this purpose.

\subsection{sMMALA with randomized adaptive selection of $\varepsilon$\label{sub:sMMALA-with-randomized}}

In this work, I propose to select the proposal step size as a function
of the current state $\mathbf{x}_{t}$ and an easily simulated auxiliary
random variable $\mathbf{w}_{t}\sim\pi(\mathbf{w})$, namely $\varepsilon=\varepsilon(\mathbf{x}_{t},\mathbf{w}_{t})$.
The role of $\mathbf{w}$ is essentially to randomize the step size,
and a further motivation for including it will be clear from section
\ref{sub:Line-search-mechanisms} when the functional forms of $\varepsilon(\cdot,\cdot)$
advocated here are discussed. Before that, I introduce the an overarching
sampling algorithm in the form of a Metropolis within Gibbs sampler
based on the sMMALA update where invariant distribution will be shown
to be the augmented target 
\begin{equation}
\pi(\mathbf{x},\mathbf{w})=\pi(\mathbf{x})\pi(\mathbf{w}).\label{eq:augmented_target}
\end{equation}
Notice that the algorithm with target (\ref{eq:augmented_target})
is primarily a construct to make the method time-homogenous and thereby
enabling easy establishment of convergence results when an adaptive
step size selection $\varepsilon(\mathbf{x},\mathbf{w})$ is employed.
In practice only the $\mathbf{x}$-subchain is of interest.

Given the current configuration $(\mathbf{x}_{t},\mathbf{w}_{t})\sim\pi(\mathbf{x},\mathbf{w})$,
one step of the proposed Metropolis within Gibbs sampler \citep[see e.g.][Algorithm A.43]{robert_casella}
based on sMMALA for the augmented target may then be summarized by
the steps
\begin{enumerate}
\item Compute forward step size $\varepsilon_{f}=\varepsilon(\mathbf{x}_{t},\mathbf{w}_{t})$.
\item Draw proposal $\mathbf{x}_{t+1}^{*}\sim N(\mathbf{x}_{t}+\frac{\varepsilon_{f}^{2}}{2}G^{-1}(\mathbf{x}_{t})g(\mathbf{x}_{t}),\varepsilon_{f}^{2}G^{-1}(\mathbf{x}_{t}))$.
\item Compute backward step size $\varepsilon_{b}=\varepsilon(\mathbf{x}_{t+1}^{*},\mathbf{w}_{t}).$
\item Compute MH accept probability 
\[
\alpha(\mathbf{x}_{t},\mathbf{x}_{t+1}^{*})=\min\left(1,\frac{\tilde{\pi}(\mathbf{x}_{t+1}^{*})\mathcal{N}\left(\mathbf{x}_{t}|\mathbf{x}_{t+1}^{*}+\frac{\varepsilon_{b}^{2}}{2}G^{-1}(\mathbf{x}_{t+1}^{*})g(\mathbf{x}_{t+1}^{*}),\varepsilon_{b}^{2}G^{-1}(\mathbf{x}_{t+1}^{*})\right)}{\tilde{\pi}(\mathbf{x}_{t})\mathcal{N}\left(\mathbf{x}_{t+1}^{*}|\mathbf{x}_{t}+\frac{\varepsilon_{f}^{2}}{2}G^{-1}(\mathbf{x}_{t})g(\mathbf{x}_{t}),\varepsilon_{f}^{2}G^{-1}(\mathbf{x}_{t})\right)}\right).
\]
Let $\mathbf{x}_{t+1}=\mathbf{x}_{t+1}^{*}$ with probability $\alpha(\mathbf{x}_{t},\mathbf{x}_{t+1}^{*})$
and let $\mathbf{x}_{t+1}=\mathbf{x}_{t}$ with remaining probability.
\item Draw $\mathbf{w}_{t+1}\sim\pi(\mathbf{w})$.
\end{enumerate}
Notice in particular that $\varepsilon_{f},\varepsilon_{b}$ are computed
using the same $\mathbf{w}$-argument and therefore steps 1-4 constitute
a reversible MH step for each $\mathbf{w}$. Based on this observation,
I prove following result: \\
\textbf{Proposition 1: }\emph{Provided that $0<\varepsilon(\mathbf{x},\mathbf{w})<\infty$
for each $(\mathbf{x},\mathbf{w})$ in the support of $\pi(\mathbf{x},\mathbf{w})$,
the Metropolis within Gibbs sampler in steps 1-5 is $\pi(\mathbf{x},\mathbf{w})$-irreducible,
aperiodic and has $\pi(\mathbf{x},\mathbf{w})$ as invariant distribution.}

The proof is given in Appendix \ref{sec:Proof-of-Proposition} and
relies on the fact that for a Metropolis within Gibbs sampler, the
parameters of the proposal distribution in the MH step for updating
one block ($\mathbf{x}$) may depend on the current state of the remaining
blocks ($\mathbf{w}$) \citep[see e.g.][]{gilks_adaptive_dir,robert_casella,zhang11quasi}.
Note in particular that steps 1-5 do not fulfill detailed balance
for target $\pi(\mathbf{x},\mathbf{w})$. On the other hand, a (time-inhomogenous
due to changing $\mathbf{w}$) detailed balance holds for each transition
of the $\mathbf{x}$-subchain. Having established that such adaptive
selection of step size does not disturb the invariant distribution
of the $\mathbf{x}$-subchain for a large class of functions $\varepsilon(\cdot,\cdot)$,
I now consider the specific functional forms of $\varepsilon(\cdot,\cdot)$
I advocate.

\subsection{Adaptive step size selection based on Hamiltonian dynamics\label{sub:Line-search-mechanisms}}

The method for selection of $\varepsilon$ uses a well-known duality
between MALA and HMC, namely that the proposal of MALA corresponds
to a single time-integration step of Euclidian metric HMC starting
at $\mathbf{x}_{t}$ when the leap frog integrator \citep{Leimkuhler:2004}
is employed \citep{1206.1901}. Here I propose to choose the step
size for sMMALA so that the energy error of a single trial step of
the corresponding (Euclidian metric) HMC method with mass matrix $G(\mathbf{x}_{t})$
is below a tunable threshold. The rationale for such an approach is
that the absolute energy error provides a measure of time integration
error that is comparable across different areas of the target density,
and that will be large if the local scaling properties of $\log\pi(\mathbf{x})$
are poorly reflected by $G(\mathbf{x})$, as in the student $t$-pilot
example discussed above. Computing the absolute energy error of the
trial step is inexpensive relative to e.g. doing complete trial sMMALA
steps for different values of $\varepsilon$, with the computational
cost typically dominated by a few additional gradient evaluations. 

To implement such energy error based selection $\mathbf{\varepsilon}(\cdot,\cdot)$,
I first take $\pi(\mathbf{w})$ to be a $d$-dimensional standard
Gaussian variable. Given the current configuration of $(\mathbf{x}_{t},\mathbf{w}_{t})$,
I define the position specific dummy Hamiltonian
\begin{equation}
\mathcal{H}(\mathbf{q}(\tau),\mathbf{p}(\tau)|\mathbf{x}_{t})=-\log\tilde{\pi}(\mathbf{q}(\tau))+\frac{1}{2}\mathbf{p(\tau)}^{T}G(\mathbf{x}_{t})^{-1}\mathbf{p}(\tau),\label{eq:dummy_hamiltonian}
\end{equation}
where $\mathbf{q}(\tau)$ denotes position and $\mathbf{p}(\tau)$
denotes momentum at fictional time $\tau$. Then a single leap frog
step of (time-) size $\varepsilon$ starting from $\mathbf{q}(0)=\mathbf{x}_{t}$
and $\mathbf{p}(0)=L(\mathbf{x}_{t})\mathbf{w}_{t}$ is performed:
\begin{eqnarray}
\mathbf{p}(\varepsilon/2) & = & \mathbf{p}(0)+\frac{\varepsilon}{2}g(\mathbf{q}(0))=L(\mathbf{x}_{t})\mathbf{w}_{t}+\frac{\varepsilon}{2}g(\mathbf{x}_{t}),\nonumber \\
\mathbf{q(\varepsilon)} & = & \mathbf{q}(0)+\varepsilon G(\mathbf{x}_{t})^{-1}\mathbf{p}(\varepsilon/2)=\mathbf{x}_{t}+\frac{\varepsilon^{2}}{2}G^{-1}(\mathbf{x}_{t})g(\mathbf{x}_{t})+\varepsilon L(\mathbf{x}_{t})^{-T}\mathbf{w}_{t}=\mathbf{x}^{*}(\varepsilon|\mathbf{x}_{t},\mathbf{w}_{t}),\label{eq:proposal_eq}\\
\mathbf{p}(\varepsilon) & = & \mathbf{p}(\varepsilon/2)+\frac{\varepsilon}{2}g(\mathbf{q}(\varepsilon))=L(\mathbf{x}_{t})\mathbf{w}_{t}+\frac{\varepsilon}{2}\left(g(\mathbf{x}_{t})+g(\mathbf{x}^{*}(\varepsilon|\mathbf{x}_{t},\mathbf{w}_{t}))\right).\nonumber 
\end{eqnarray}
The trial proposal $\mathbf{x}^{*}(\varepsilon|\mathbf{x}_{t},\mathbf{w}_{t})$
would occur if the standard normal vector $\mathbf{w}_{t}$ was used
to generate a proposal from (\ref{eq:sMMALA_prop_distr}) with current
state equal to $\mathbf{x}_{t}$ and time step size $\varepsilon$.
The energy error associated with this trial time integration step
is given as 
\begin{eqnarray*}
\Delta(\varepsilon|\mathbf{x}_{t},\mathbf{w}_{t}) & = & \mathcal{H}(\text{\ensuremath{\mathbf{q}}}(0),\mathbf{p}(0)|\text{\ensuremath{\mathbf{x}}}_{t})-\mathcal{H}(\mathbf{q}(\varepsilon),\mathbf{p(\varepsilon)}|\mathbf{x}_{t})\\
 & = & -\log\tilde{\pi}(\mathbf{x}_{t})+\log\tilde{\pi}(\mathbf{x}^{*}(\varepsilon|\mathbf{x}_{t},\mathbf{w}_{t}))\\
 &  & -\frac{\varepsilon}{2}\mathbf{w}_{t}^{T}\mathbf{r}-\frac{\varepsilon^{2}}{8}\mathbf{r}^{T}\mathbf{r}
\end{eqnarray*}
where $\mathbf{r}=L(\mathbf{x}_{t})^{-1}\left(g(\mathbf{x}_{t})+g(\mathbf{x}^{*}(\varepsilon|\mathbf{x}_{t},\mathbf{w}_{t}))\right)$.
Based on the expression for $\Delta(\varepsilon|\mathbf{x}_{t},\mathbf{w}_{t})$,
a possible method for choosing $\varepsilon=\varepsilon(\mathbf{x},\mathbf{w}$)
would involve the following steps: 

Suppose $\gamma>0$ is the tunable maximal allowable trial energy
error, with lower values of $\gamma$ corresponding to higher acceptance
rates and smaller $\varepsilon$. Let $0<\bar{\varepsilon}<\infty$
be the maximal step size and let $\varepsilon_{0}$ denote the smallest
positive root in $\varepsilon$ of $|\Delta(\varepsilon|\mathbf{x},\mathbf{w})|=\gamma$.
An idealized candidate for choosing $\varepsilon$ is then $\varepsilon(\mathbf{x},\mathbf{w})=\min(\bar{\varepsilon},\varepsilon_{0})$.
The $\varepsilon>0$ bound required in Proposition 1 is automatically
fulfilled as consequence smoothness assumptions on $\log\tilde{\pi}(\mathbf{x})$,
and thus such a step size function would lead the sampling algorithm
outlined in section \ref{sub:sMMALA-with-randomized} to have the
correct invariant distribution. In practice, locating $\varepsilon_{0}$
would amount to solving an equation that involves the gradient of
the log-target numerically, and for computational effectivity reasons
I therefore propose a method for only approximating $\varepsilon_{0}$
in Section \ref{sub:Practical-implementation-and}. 

It is worth noticing that the Hamiltonian (\ref{eq:dummy_hamiltonian})
should be interpreted in the Euclidian metric sense as a tool to measure
the time integration error around $\mathbf{q}(0)=\mathbf{x}_{t}$,
$\mathbf{p}(0)=L(\mathbf{x}_{t})\mathbf{w}_{t}$ when $G(\mathbf{x}_{t})$
is the fixed mass matrix. This is relevant for sMMALA as the distribution
of proposed $\mathbf{q}(\varepsilon)$-states is (\ref{eq:sMMALA_prop_distr})
in this case. Moreover, it is in contrast to interpreting (\ref{eq:dummy_hamiltonian})
in a Riemann manifold HMC sense with Hamiltonian say $-\log\tilde{\pi}(\mathbf{q})+\frac{1}{2}\mathbf{p}^{T}G(\mathbf{q})^{-1}\mathbf{p}$
(which is off target due to a missing $\frac{1}{2}\log|G(\mathbf{q})|$
term \citep{girolami_calderhead_11,1212.4693}). Under the latter
interpretation, the energy error is irrelevant for sMMALA as the proposed
$\mathbf{q}(\varepsilon)$ will not be distributed according to (\ref{eq:sMMALA_prop_distr})
regardless of the numerical integrator being employed, as the time
evolution of $\mathbf{p}$ will involve the gradient of $\frac{1}{2}\mathbf{p}^{T}G(\mathbf{q})^{-1}\mathbf{p}$
with respect to $\mathbf{q}$. In the same vein, I notice that $\min(1,\exp(\Delta))$
cannot be interpreted as a trial acceptance probability for the sMMALA
as the sMMALA deviates from being a reversible and volume-preserving
discretization of either Hamiltonian when the $\mathbf{x}_{t}$-dependence
of the mass matrix is taken into account in (\ref{eq:dummy_hamiltonian}).
Rather, $\gamma$ should be interpreted as a tolerance of local (irreversible)
time integration energy error, and in particular its relationship
to the acceptance probability of the overall (reversible) sMMALA step
is non-linear.

The role of $\mathbf{w}$ is to act as typical (standardized) momentum
variable that changes in each iteration of the overarching MCMC procedure
to reflect the variation in the distribution of the proposal (\ref{eq:sMMALA_prop_distr}).
A further potential candidate would be to integrate out $\mathbf{w}$,
e.g. $\varepsilon(\mathbf{x})=\int\varepsilon(\mathbf{x},\mathbf{w})\pi(\mathbf{w})d\mathbf{w}$
so that steps 1-4 of section \ref{sub:sMMALA-with-randomized} would
constitute a time-homogenous and reversible MH step. However even
for moderate dimension $d$ this becomes computationally intractable,
as this integral has to be computed numerically. Further, it is not
obvious that a deterministic (conditional on $\mathbf{x}$) step size
$\varepsilon(\mathbf{x})$ has clear advantages over a stochastic
step size $\varepsilon(\mathbf{x},\mathbf{w})$.

\begin{figure}
\centering{}\includegraphics[scale=0.8]{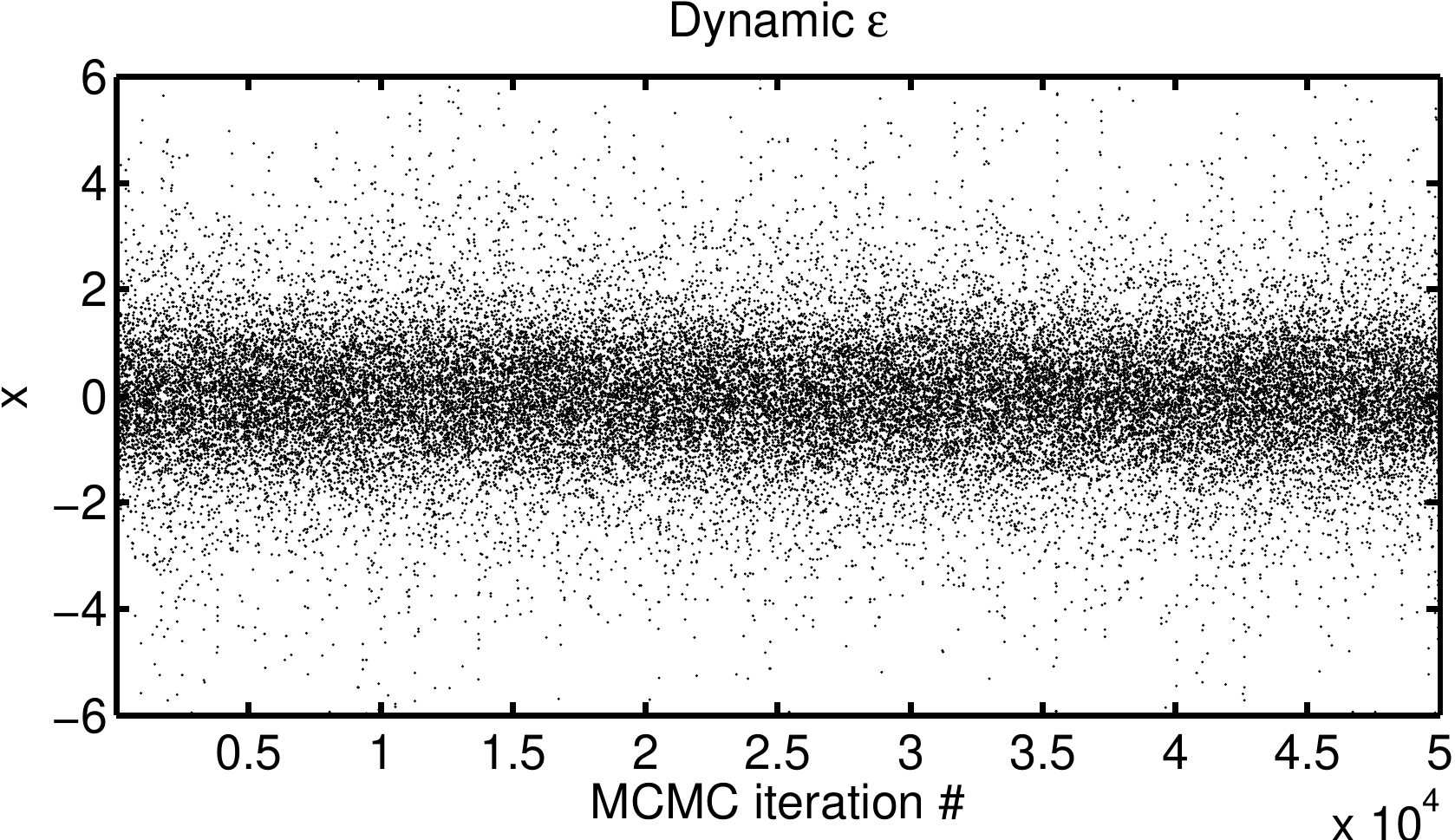}\protect\caption{\label{fig:dyn_trace-t}50,000 iterations of sMMALA with metric tensor
(\ref{eq:1D-metric}), $u=0.001$, $\varepsilon(\mathbf{x},\mathbf{w})=\min(1,\varepsilon_{0})$
as defined in the text and $\gamma=1.0$ applied to a $t$-distribution
with 4 degrees of freedom. It is seen that the adaptive selection
of step size resolves sampling in the problematic regions around the
inflection points.}
\end{figure}
\begin{figure}
\centering{}\includegraphics[scale=0.5]{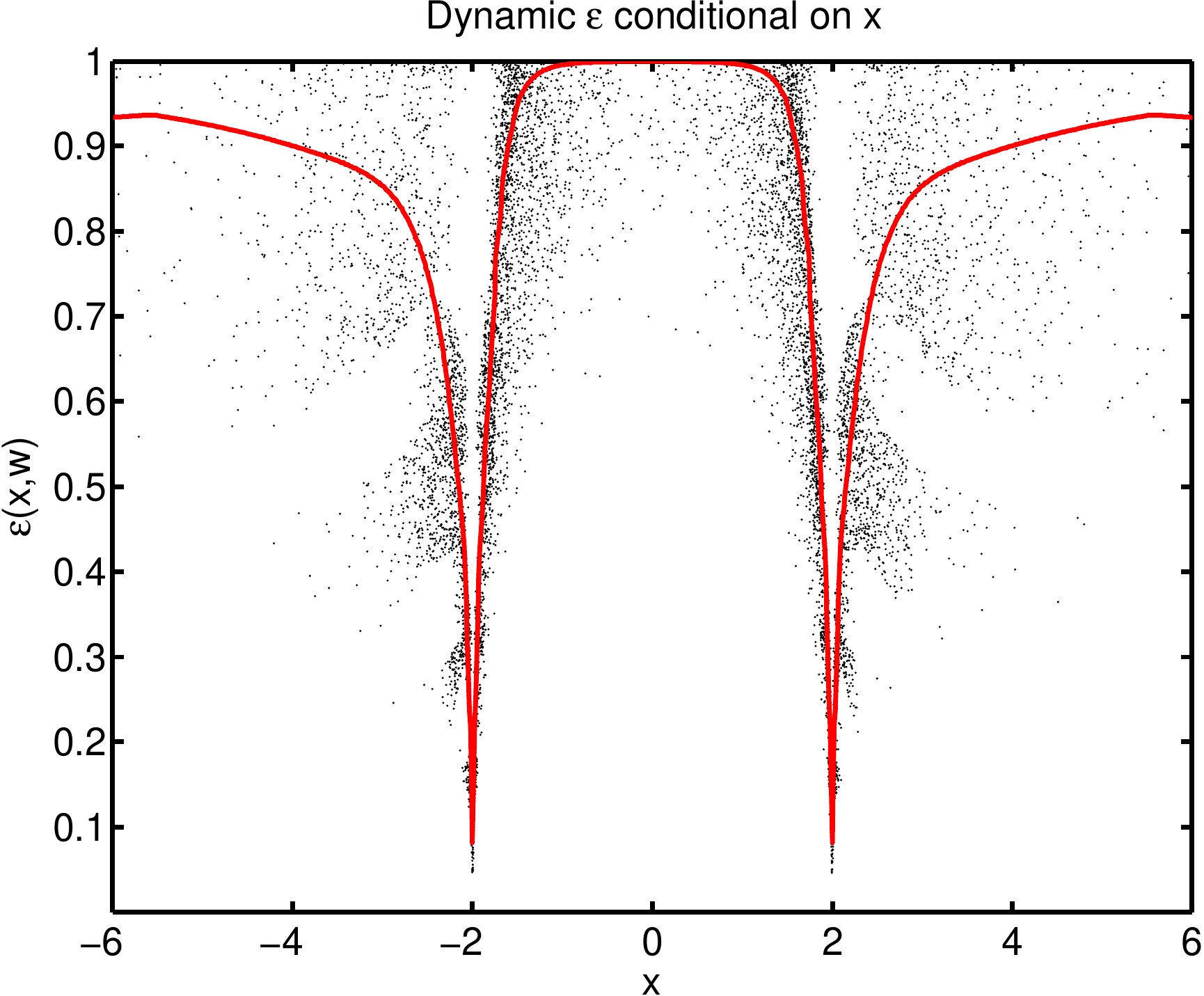}\protect\caption{\label{fig:Values-of-the}Values of the (forward) step size $\varepsilon(\mathbf{x}_{t},\mathbf{w}_{t})=\min(1,\varepsilon_{0})$
as a function of $\mathbf{x}_{t}$ for the simulation presented in
Figure \ref{fig:dyn_trace-t} are given as dots. The solid line indicates
the average adaptive step size calculated as $\int\varepsilon(\mathbf{x},\mathbf{w})p(\mathbf{w})d\mathbf{w}$
as a function of $\mathbf{x}$.}
\end{figure}

To illustrate how the adaptive step size selection $\varepsilon(\mathbf{x},\mathbf{w})=\min(1,\varepsilon_{0})$
resolves the shortcomings of sMMALA near inflection points, I consider
again the $t_{4}-$distribution target considered in section \ref{sub:The-GMW-factorization}
with metric tensor (\ref{eq:1D-metric}). Figure \ref{fig:dyn_trace-t}
presents 50,000 MCMC iterations using adaptive step size sMMALA using
a maximal absolute energy error $\gamma=1.0$. It is seen that the
adaptive step size selection resolves the inefficient sampling near
the inflection points seen for the fixed step size implementation
in Figure \ref{fig:fixed-trace-t}. Figure \ref{fig:Values-of-the}
shows the actual values of the (forward) adaptive step sizes $\varepsilon(\mathbf{x}_{t},\mathbf{w}_{t})$
as a function of $\mathbf{x}_{t}$, along with their expected value
found by integrating out $\mathbf{w}$. It is seen that the energy
error criterion appear to discriminate very well that the metric tensor
(\ref{eq:1D-metric}) shows pathological behavior around the inflection
points at $|\mathbf{x}|=2.$

As pointed out by a reviewer, the above energy error argument can
also be applied in non-adaptive step size sMMALA methods to locate
regions of the support where deficiencies in the metric tensor lead
to inefficient sampling analogous to what is seen in Figure \ref{fig:fixed-trace-t}.
This is in particularly of interest for $d\geq3$, as then regions
with inefficient sampling could be difficult to spot using e.g. trace
or scatter plots. In practice, using energy error for such an end
would amount to calculating the energy error associated with each
proposed state. Specifically, suppose $\mathbf{x}_{t}$ is the current
state of a non-adaptive sMMALA method and $\mathbf{z}\sim N(0,I_{d})$
is the standardized variable used to generate a proposal $\mathbf{x}^{*}$
according to (\ref{eq:sMMALA_prop_distr}). The forward (backward)
energy error $\Delta_{t}^{f}$ ($\Delta_{t}^{b}$) associated with
the proposal $\mathbf{x}^{*}$ are given by 
\begin{eqnarray*}
\Delta_{t}^{f} & = & -\log\tilde{\pi}(\mathbf{x}_{t})+\log\tilde{\pi}(\mathbf{x}^{*})-\frac{\varepsilon}{2}\mathbf{z}^{T}\mathbf{r}_{f}-\frac{\varepsilon^{2}}{8}\mathbf{r}_{f}^{T}\mathbf{r}_{f}\\
\Delta_{t}^{b} & = & -\log\tilde{\pi}(\mathbf{x}^{*})+\log\tilde{\pi}(\mathbf{x}_{t})-\frac{\varepsilon}{2}\mathbf{s}^{T}\mathbf{r}_{b}-\frac{\varepsilon^{2}}{8}\mathbf{r}_{b}^{T}\mathbf{r}_{b}\\
\mathbf{s} & = & \varepsilon^{-1}L(\mathbf{x}^{*})^{T}\left[\mathbf{x}_{t}-\mathbf{x}^{*}-\frac{\varepsilon^{2}}{2}G^{-1}(\mathbf{x}^{*})g(\mathbf{x}^{*})\right]
\end{eqnarray*}
where $\mathbf{r}_{f}=L(\mathbf{x}_{t})^{-1}\left(g(\mathbf{x}_{t})+g(\mathbf{x}^{*})\right)$
and $\mathbf{r}_{b}=L(\mathbf{x}^{*})^{-1}\left(g(\mathbf{x}_{t})+g(\mathbf{x}^{*})\right)$.
A large $|\Delta_{t}^{f}|$ ($|\Delta_{t}^{b}|$) (e.g. $|\Delta_{t}^{k}|>5,\; k=f,b$)
will indicate that local scaling properties in the region around $\mathbf{x}_{t}$
($\mathbf{x}^{*}$) are poorly reflected by $G(\mathbf{x}_{t})$ ($G(\mathbf{x}^{*})$)
and the user should consider revising the specification of $G$ or
at least reduce $\varepsilon$. Note in particular that $\Delta_{t}^{k},\; k=f,b$
can be computed in each iteration with minimal additional cost as
the involved log-kernels, gradients and metric tensors must be computed
anyway to form the acceptance probability for proposal $\mathbf{x}^{*}$.

\subsection{Practical implementation and choosing $\gamma$\label{sub:Practical-implementation-and}}

\begin{algorithm}
\begin{ntabbing}
\hspace{2em} \= \hspace{2em} \= \hspace{2em} \= \\
Set $\varepsilon_1\leftarrow 1$ (or some reasonable $\bar \varepsilon$)  \label{}\\ 
Set $\beta \leftarrow 10$ (or some reasonable energy error threshold). \label{}\\
Set $\rho \leftarrow 0.5$ (or some reasonable step size decrement). \label{}\\ 
{\bfseries for}($s=1,2,\dots$ ) \label{}\\
\> Compute $|\Delta(\varepsilon_{s}|\mathbf{x},\mathbf{w})|$.\label{}\\
\> {\bfseries if}($|\Delta(\varepsilon_{s}|\mathbf{x},\mathbf{w})|>\beta$)\label{}\\
\> \> $\varepsilon_{s+1} \leftarrow \rho \varepsilon_s$.\label{}\\
\> {\bfseries else} \label{}\\
\> \> {\bfseries if}($|\Delta(\varepsilon_{s}|\mathbf{x},\mathbf{w})|<\gamma$)\label{}\\
\> \> \> {\bfseries Return} $\varepsilon(\mathbf{x},\mathbf{w}) = \varepsilon_s$.\label{}\\
\> \> {\bfseries else}\label{}\\
\> \> \> $\varepsilon_{s+1} \leftarrow 0.95 \left( \frac{\gamma}{|\Delta(\varepsilon_{s}|\mathbf{x},\mathbf{w})|} \right)^{1/3} \varepsilon_{s}$.\label{}\\
\> \> {\bfseries end if}\label{}\\
\> {\bfseries end if}\label{}\\
{\bfseries end for}\label{}\\
\end{ntabbing}

\protect\caption{\label{alg:line_search}Practical backtracking line search algorithm. }
\end{algorithm}
For practical applications it is not necessary, nor desirable from
a computational perspective, to find $\varepsilon_{0}$ with high
precision. Rather a backtracking iterative scheme as exemplified in
Algorithm \ref{alg:line_search} can be used. This algorithm is inspired
by line searches commonly used numerical optimization \citep{noce:wrig:1999}
and the overarching idea is to generate a sequence trial step sizes
$\bar{\varepsilon}=\varepsilon_{1}>\varepsilon_{2}>\dots$ until the
criterion $|\Delta(\varepsilon_{s}|\mathbf{x},\mathbf{w})|<\gamma$
is fulfilled. Algorithm \ref{alg:line_search} has two types of decrements,
where in the case that absolute energy error is greater than $\beta$,
the next trial step size is a factor $\rho<1$ times the incumbent
one. If the absolute energy error is smaller than $\beta$, but greater
than $\gamma$, the choice of the next trial step size is informed
by the fact that $\Delta(\varepsilon|\mathbf{x},\mathbf{w})=O(\varepsilon^{3})$
for small $\varepsilon$. More specifically, I let next trial step
$\varepsilon_{s+1}$ be 0.95 times the root in $\varepsilon$ of $\hat{\Delta}_{s}(\varepsilon)-\gamma$
where $\hat{\Delta}_{s}(\varepsilon)=(\varepsilon/\varepsilon_{s})^{3}|\Delta(\varepsilon_{s}|\mathbf{x},\mathbf{w})|$
is the third order monomial that interpolates the observed absolute
energy error at $\varepsilon_{s}$. The factor 0.95 is included to
counteract slow convergence infrequently seen when $|\Delta(\varepsilon_{s}|\mathbf{x},\mathbf{w})|$
is close to $\gamma$. Of course Algorithm \ref{alg:line_search}
constitute only an example of how to implement $\varepsilon(\mathbf{x},\mathbf{w})$,
and may be further tuned and refined for each model instance without
disturbing the stationary distribution as long as it remains the same
throughout the MCMC simulation and fulfills the assumptions in Proposition
1. The cost of each iteration will typically be dominated by the gradient
evaluation needed for computing $\mathbf{r}$ for each trial $\varepsilon$. 

For non-Gaussian targets, $\gamma$ needs to be tuned to obtain e.g.
the desired acceptance rate or to maximize some more direct performance
measure such as the expected jumping distance or effective sample
size. The tuning could be done by e.g. dual averaging during the burn
in iterations as in \citet{JMLR:v15:hoffman14a}. However, I have
found that as a rule of thumb, values of $\gamma$ between 1 and 2
tends to produce acceptable results for low to moderate-dimensional
problems as the ones considered below. In this case, the adaptive
step size selection typically produces long step sizes ($\varepsilon\sim\bar{\varepsilon}$)
when $G$ contains useful scaling information, but acts as safeguard
and reduces the step size substantially when $G$ shows some form
of pathological behavior. 

It is interesting to consider the impact of dimension $d$ on the
proposed adaptive step size selection, and consequently on the overall
performance of the proposed methodology. For this purpose, I consider
any $d$-dimensional non-degenerate $N(\mu,\Sigma)$ target, as it
is rather straightforward to show that 
\[
E_{(\mathbf{x},\mathbf{w})}\left(\Delta(\varepsilon|\mathbf{x},\mathbf{w})\right)=d\left(\frac{1}{4}\varepsilon^{4}-\frac{1}{32}\varepsilon^{6}\right),
\]
when $G(\mathbf{x})=-H(\mathbf{x})=\Sigma^{-1}$. For any fixed $\gamma$,
which in the Gaussian case translates to a fixed acceptance rate invariant
of $d$, the adaptive step size has leading term behavior $\varepsilon=O(d^{-1/4})$.
Thus for high-dimensional target distributions, it is inevitable that
adaptive step size sMMALA will revert to random walk behavior, and
as a consequence I consider primarily low to moderate-dimensional
applications as the primary scope.

\section{Illustrations}

This section considers example models and compares the proposed Adaptive
step size Modified Hessian MALA (AMH-MALA) methodology to alternatives.
The GARCH(1,1) model with $t$-distributed innovations considered
first is included to highlight the behavior of AMH-MALA for a posterior
distribution that exhibit strong non-linear dependence structures
and has indefinite Hessian in substantial parts of the relevant parameter
space. The Bayesian binary response regressions considered afterwards
are included to allow for easy comparison with the methodology presented
in \citet{girolami_calderhead_11}. 

To compare the performance of different MCMC methods, I follow \citet{girolami_calderhead_11}
in estimating effective sample size (ESS) via the initial monotone
sequence estimator of \citet{geyer1992}, and in particular compare
the minimum ESS (across the $d$ dimensions of $\pi$) per second
CPU-time spent obtaining the samples. All computations were carried
out on a 2014 macbook pro with a 2 GHz Intel Core i7 processor and
8 GB of memory. The code used for the simulations is available at
http://www.ux.uis.no/\textasciitilde{}tore/code/adaptive\_langevin/,
including a C++ function implementing the dense modified Cholesky
factorization that can be called from Matlab.

\subsection{GARCH(1,1) model with $t$-distributed innovations}

\begin{table}
\begin{centering}
\begin{tabular}{lcccccc}
\hline 
Method & CPU time & $\alpha_{0}$ & $\alpha_{1}$ & $\beta$ & $\nu$ & minimum ESS\tabularnewline
 & (s) & ESS & ESS & ESS & ESS & per second\tabularnewline
\hline 
bayesGARCH & 18.7 & 80.0 & 52.9 & 48.2 & 63.9 & \textbf{(2.46)}\tabularnewline
AMH-MALA & 140 & 283 & 310 & 252 & 398 & \textbf{1.79}\tabularnewline
AMH-MALA(eig) & 146 & 251 & 262 & 229 & 424 & 1.51\tabularnewline
HMC & 2653 & \textbf{4738} & \textbf{2284} & \textbf{2503} & \textbf{4993} & 0.86\tabularnewline
\hline 
\end{tabular}\protect\caption{\label{tab:GARCH-Effective-samples-sizes}Effective samples sizes
and minimum effective samples sizes per second CPU time for different
MCMC algorithms applied to the GARCH(1,1) model with $t$-distributed
innovations. Best performances are indicated in bold font. All figures
are calculated over 5000 MCMC iterations after burn in. Note that
the bayesGARCH Gibbs sampler is partially written in C, and thus the
CPU-time and minimum ESS per CPU time are not directly comparable
to the remaining figures. }

\par\end{centering}

\end{table}
\begin{figure}
\centering{}\includegraphics[scale=0.9]{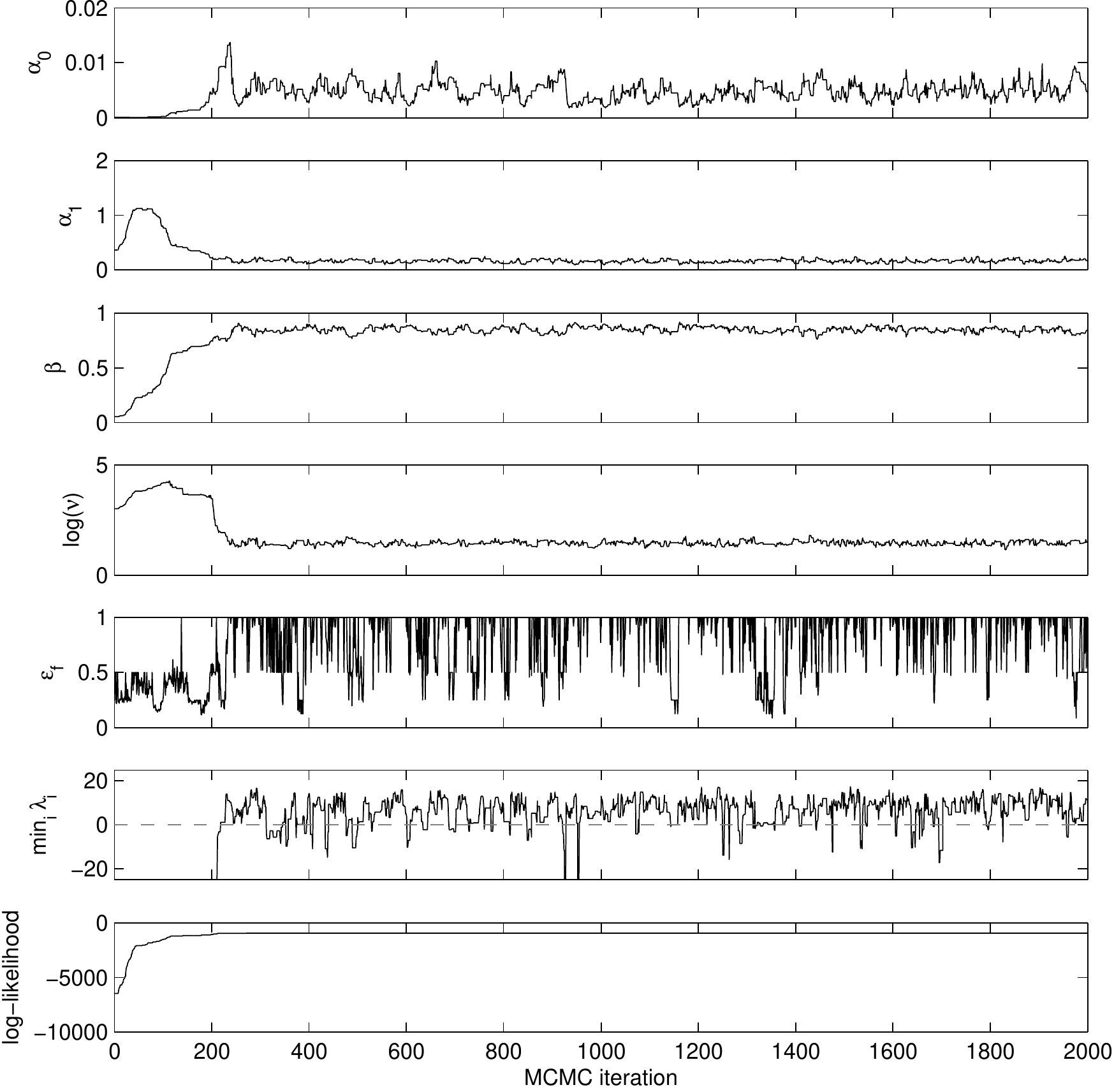}\protect\caption{\label{fig:garch-Diagnostics}Diagnostics for a typical run of AMH-MALA
with initial parameters set to the unrealistic values $\log(\alpha_{0})=-10$,
$\log(\alpha_{1})=-1$, $\log(\beta)=-3$ and $\nu=20$. The upper
4 panels are trace plots of parameters $\alpha_{0},\alpha_{1},\beta$
and $\log(\nu)$ where the logarithm is applied to latter for visual
reasons. The 5. panel is a trace plot of the forward step size $\varepsilon_{f}$.
The 6. panel depicts the smallest eigenvalue of $-H(\mathbf{x}_{t})$,
with smallest eigenvalue for the first 211 iterations being smaller
than -25. The last panel shows the log-target target density at the
current iteration, i.e. $\log\tilde{\pi}(\mathbf{x}_{t})$.}
\end{figure}
As the first realistic illustration I consider sampling the posterior
parameters of a GARCH(1,1) model with $t$-distributed innovations
\citep[see e.g.][Chapter 4]{QRM_mcneil} for log-return observations
$y_{1},\dots,y_{T}$ on the form
\begin{eqnarray}
y_{i} & = & \sqrt{h_{i}}\eta_{i},\;\eta_{i}=\sqrt{\frac{\nu-2}{\nu}}\tilde{\eta}_{i},\;\tilde{\eta}_{i}\sim\text{iid }t(\nu),\; i=1,\dots,T,\label{eq:garch1}\\
h_{i} & = & \alpha_{0}+\alpha_{1}y_{i-1}^{2}+\beta h_{i-1},\; i=2,\dots,T,\; h_{1}=\alpha_{0}.\label{eq:garch2}
\end{eqnarray}
The priors are taken from \citet{RJournal_2010-2_Ardia+Hoogerheide}
and are the default priors used in the associated \texttt{bayesGARCH}
R-package for the same model. Specifically, $\alpha_{0},\alpha_{1},\beta$
have independent truncated normal priors $p(\alpha_{k})\propto\exp(-\frac{1}{2}\frac{\alpha_{k}^{2}}{1000})\mathbf{1}_{\left\{ \alpha_{k}>0\right\} },\; k=0,1$
and $p(\beta)\propto\exp(-\frac{1}{2}\frac{\beta^{2}}{1000})\mathbf{1}_{\left\{ \beta>0\right\} }$.
The prior for the degrees of freedom parameter $\nu$ is a truncated
exponential, namely $p(\nu)\propto\exp(-\nu/100)\mathbf{1}_{\left\{ \nu>2\right\} }$.
For our illustration, I employ a data set of Deutschmark vs British
Pound (DEM/GBP) foreign exchange rate log-returns at daily frequency,
which is also included in the \texttt{bayesGARCH} package. The sample
covers January 3, 1985 to December 31, 1991, and constitute $T=1974$
log-return observations. 

The proposed AMH-MALA methodology is compared to the Gibbs sampler
implemented in the \texttt{bayesGARCH} package \citep{RJournal_2010-2_Ardia+Hoogerheide}
and HMC with mass matrix fixed to the identity matrix. For HMC I follow
\citet{girolami_calderhead_11} and choose 100 time integration steps
and the integration step size was taken to be $\varepsilon=0.0075\pm10\%$
uniform noise to attain acceptance rates of around 80\%. In addition
I also consider AMH-MALA implemented with a metric tensor similar
to the one proposed by \citet{1212.4693} and denote this by AMH-MALA(eig).
Specifically, let $\lambda_{i}$ denote the $i$th eigenvalue of $-H$.
Then $G$ has the same eigenvectors as $-H$ but the $i$th eigenvalue
is taken as $\max(|\lambda_{i}|,0.001)$. For both AMH-MALA methods,
I use $\gamma=1.0$ and Algorithm \ref{alg:line_search} with $\beta=10.0$,
$\rho=0.5$ for adaptive step size selection. For AMH-MALA based on
the modified Cholesky, I use $u=0.001$. HMC and the AMH-MALA methods
are applied in $\log$-transformed parameters $\alpha_{0}^{\prime}=\log(\alpha_{0})$,
$\alpha_{1}^{\prime}=\log(\alpha_{1})$, $\beta^{\prime}=\log(\beta)$
and $\nu^{\prime}=\log(\nu-2)$ to remove potential numerical problems
related to the truncations imposed by the priors.

Table \ref{tab:GARCH-Effective-samples-sizes} provides mean CPU times,
mean ESSes and mean minimum ESS per computing time for the parameters
of the GARCH model (\ref{eq:garch1}-\ref{eq:garch2}) across 10 independent
repeated runs of 5000 samples each. For \texttt{bayesGARCH} I used
5000 burn in iterations, whereas the remaining methods used 1000 burn
in iterations. The reported CPU times excludes burn in. It should
be noted that the Gibbs sampler in the \texttt{bayesGARCH} package
is written partially in C and the computing times are therefore not
directly comparable to the remaining methods, which are implemented
fully in Matlab. It is seen that AMH-MALA produces the most effective
samples per unit computing time for the methods written in Matlab,
and also produces substantially better ESSes per iteration than the
\texttt{bayesGARCH}. AMH-MALA and AMH-MALA(eig) produces similar results,
indicating that there is little added value to using full spectral
decomposition over the modified Cholesky factorization for this situation.

Figure \ref{fig:garch-Diagnostics} depicts various diagnostic output
for a typical run of AMH-MALA applied to the parameters of (\ref{eq:garch1}-\ref{eq:garch2}).
The initial parameters are set to highly ill informed values to illustrate
the different behavior of AMH-MALA in transient and stationary regimes.
It is seen that AMH-MALA takes shorter forward steps in the transient
regime up to approximately iteration 220. The negative Hessian is
severely indefinite for these iterations. This shows that choosing
step size based on the energy error enables to AMH-MALA to make progress
in the transient regime with the same tuning parameters as are used
in the stationary regime. To contrast this, fixing $\varepsilon_{f}=\varepsilon_{b}=1$
but otherwise keeping the setup identical to that reported in Figure
\ref{fig:garch-Diagnostics} results in a single accepted move in
the course of the 2000 iterations, indicating the need for different
tuning parameters in stationary and transient regimes \citep{RSSB:RSSB500}. 

Visual inspection of Figure \ref{fig:garch-Diagnostics} indicates
that small values of $\varepsilon_{f}$ are often associated with
near zero minimum eigenvalues. This can be statistically confirmed
as the correlation between $\log|\min_{i}\lambda_{i}|$ and $\varepsilon_{f}$
is 0.63 for iterations 220-2000. This indicates that the adaptive
step size procedure on average sees both large positive- and large
negative curvature information as useful and consequently takes long
step sizes, whereas it takes shorter step sizes when very little curvature
information is available in some direction. Still there is a small
positive correlation of 0.24 between $\min_{i}\lambda_{i}$ and the
acceptance probability, indicating that AMH-MALA performs slightly
better in regions of the support where the target is log-concave.

\subsection{Bayesian binary response regression}

\begin{table}
\begin{centering}
\begin{tabular}{llccccc}
\hline 
Model & Method & CPU time  & minimum & median & maximum & minimum ESS\tabularnewline
 &  & (s) & ESS & ESS & ESS & per second\tabularnewline
\hline 
\multicolumn{7}{c}{Australian credit data set ($d=15$, $n=690$)}\tabularnewline
\hline 
logit & AMH-MALA & 2.6 & \textbf{ 436} &  579 &  725 &  167\tabularnewline
logit & sMMALA & 1.6 & \textbf{ 436} & \textbf{ 591} & \textbf{ 745} &  \textbf{271}\tabularnewline
\hline 
probit & AMH-MALA & 3.4 & \textbf{ 575} & \textbf{ 774} & \textbf{ 919} &  167\tabularnewline
probit & sMMALA & 1.8 &  491 &  665 &  869 & \textbf{ 280}\tabularnewline
probit & Adaptive sMMALA & 3.3 &  475 &  652 &  809 &  142\tabularnewline
\hline 
\multicolumn{7}{c}{German credit data set ($d=25,$ $n=1000$)}\tabularnewline
\hline 
logit & AMH-MALA & 3.6 &  417 &  570 &  707 &  117\tabularnewline
logit & sMMALA & 2.1 & \textbf{ 432} & \textbf{ 616} & \textbf{ 751} & \textbf{ 202}\tabularnewline
\hline 
probit & AMH-MALA &  4.5 & \textbf{ 579} & \textbf{ 747} & \textbf{ 886} &  129\tabularnewline
probit & sMMALA &  2.4 &  534 &  700 &  846 & \textbf{ 226}\tabularnewline
probit & Adaptive sMMALA &  4.4 &  488 &  649 &  785 &  111\tabularnewline
\hline 
\multicolumn{7}{c}{Heart data set ($d=14,$ $n=270$)}\tabularnewline
\hline 
logit & AMH-MALA & 2.0 &  362 &  468 &  579 &  176\tabularnewline
logit & sMMALA & 1.3 & \textbf{ 364} & \textbf{ 473} & \textbf{ 587} & \textbf{ 273}\tabularnewline
\hline 
probit & AMH-MALA &  2.5 & \textbf{ 612} & \textbf{ 760} & \textbf{ 883} &  247\tabularnewline
probit & sMMALA & 1.4 &  451 &  593 &  734 &  \textbf{331}\tabularnewline
probit & Adaptive sMMALA & 2.4 &  475 &  590 &  719 &  197\tabularnewline
\hline 
\multicolumn{7}{c}{Pima Indian data set ($d=8$, $n=532$)}\tabularnewline
\hline 
logit & AMH-MALA &  2.2 & \textbf{ 1043} & \textbf{ 1184} &  1296 &  479\tabularnewline
logit & sMMALA &  1.4 &  1010 &  1174 & \textbf{ 1314} & \textbf{ 732}\tabularnewline
\hline 
probit & AMH-MALA &  2.7 & \textbf{ 1143} & \textbf{ 1282} &  1428 &  425\tabularnewline
probit & sMMALA & 1.4 &  1089 &  1253 & \textbf{ 1480} & \textbf{ 752}\tabularnewline
probit & Adaptive sMMALA &  2.6 &  1116 &  1242 &  1454 &  431\tabularnewline
\hline 
\multicolumn{7}{c}{Ripley data set ($d=7$, $n=250$)}\tabularnewline
\hline 
logit & AMH-MALA &  1.3 &  285 &  375 &  \textbf{467} &  227\tabularnewline
logit & sMMALA &  0.6 & \textbf{ 291} & \textbf{ 387} &  464 & \textbf{ 506}\tabularnewline
\hline 
probit & AMH-MALA & 1.8 & \textbf{ 387} & \textbf{ 536} & \textbf{ 623} &  212\tabularnewline
probit & sMMALA &  0.8 &  287 &  379 &  490 & \textbf{ 381}\tabularnewline
probit & Adaptive sMMALA & 1.8 &  293 &  391 &  506 &  161\tabularnewline
\hline 
\end{tabular}
\par\end{centering}

\protect\caption{\label{tab:Effective-sample-sizes-binary}Effective sample sizes and
CPU times for the Bayesian binary response regressions. Best performances
are indicated with bold font. ``logit'' correspond models with logistic
link function. In this case the negative Hessian and Fisher information
matrix coincides. ``probit'' correspond to the standard normal cumulative
distribution function as the inverse link function. In this case the
negative Hessian is different from the Fisher information matrix.
sMMALA replicates the simplified manifold MALA of \citet{girolami_calderhead_11}
whereas Adaptive sMMALA uses the metric tensor of \citet{girolami_calderhead_11}
along with adaptive step size selection.}
\end{table}
This section considers Bayesian inference for two types of binary
response generalized linear models. The models are included in order
to compare the proposed methodology with that of \citet{girolami_calderhead_11}.
Specifically, I consider models for observed binary responses $\mathbf{y}=(y_{1},\dots,y_{n})$
on the form
\begin{eqnarray}
P(y_{i} & = & 1)=\rho\left[(X\beta)_{i}\right],\; i=1,\dots,n,\label{eq:bin_reg_obs}\\
\beta & \sim & N(0,100I_{d}).\nonumber 
\end{eqnarray}
Here $X\in\mathbb{R}^{n\times d}$ is a design matrix and the inverse
link function $\rho$ is specified either as a logit link corresponding
to $\rho(x)=\exp(x)/(1+\exp(x))$ or the probit link corresponding
to $\rho(x)=\Phi(x)$ where $\Phi$ denotes the $N(0,1)$ distribution
function. For both link functions, the Fisher information matrix is
available in closed form, and for the logit link the negative Hessian
of log-likelihood function associated with (\ref{eq:bin_reg_obs})
coincides with the Fisher information matrix, whereas this is not
the case for the probit model. However, the negative Hessian is still
positive definite in the relevant region for the probit model.

I consider the collection of 5 data sets used by \citet{girolami_calderhead_11}
where $n$ ranges between 250 and 1000 and $d$ ranges between 7 and
25. The sMMALA method using $G_{GC}$ is used as a reference. For
the logit model, this amounts to a replication of the simplified MMALA-part
of the Monte Carlo experiment of \citet{girolami_calderhead_11},
and therefore admits calibration of their results against those presented
here. In particular, I recoded the method of \citet{girolami_calderhead_11}
so that all codes use the same routines for evaluating likelihoods,
gradients and Hessians to make the comparison as fair as possible.
For AMH-MALA I employed $\gamma=2.0$ and Algorithm \ref{alg:line_search}
with $\beta=20.0$, $\rho=0.7$. In this setting, Algorithm \ref{alg:line_search}
works mainly as a safeguard against numerical problems occurring when
some $\rho\left[(X\beta)_{i}\right]\rightarrow0$ or $\rho\left[(X\beta)_{i}\right]\rightarrow1$.
The results are presented in Table \ref{tab:Effective-sample-sizes-binary}.
Through out, I collect 5000 samples after 5000 iterations of burn
in, and the timings are for producing the post burn in samples. All
experiments are repeated 10 times and reported numbers are averages
across these replica. In \citet{girolami_calderhead_11}, the simplified
MMALA was found to be the best method for the logit model for 4 out
5 data sets when other contenders included the Riemann manifold HMC
and full manifold MALA and the performance measure was the minimum
(over $\beta$) ESS produced per unit of time.

For the logit model, AMH-MALA may be interpreted as an adaptive step
size version of \citet{girolami_calderhead_11}'s simplified MMALA,
I see that the line search mechanism leads to approximately a factor
2 increase computing time, whereas the number of effective samples
produced per iteration are approximately the same. Looking at the
results reported in Table 3-7 in \citet{girolami_calderhead_11} using
simplified MMALA as a reference, I find that AMH-MALA performance
roughly on par with Riemann manifold HMC for this model and these
data sets.

For the probit model, where $-H(\beta)$ and $G_{GC}$ do not coincide,
I see slightly larger differences in terms of effective sample size
in favor of AMH-MALA, whereas the relative consumption of CPU time
is roughly the same as in the logit case. To further investigate this
find, I also implemented a method based on $G_{GC}$ but otherwise
identical to AMH-MALA and denote this method Adaptive sMMALA. It is
seen that the improved ESSes appear to be a feature of the application
of the Hessian rather than adaptive step size selection as the ESSes
of sMMALA and Adaptive sMMALA are roughly the same. From this I may
conclude that the negative Hessian may be a better alternative than
the Fisher information-based metric tensor for models where the Hessian
is positive definite in all relevant regions.

\section{Discussion }

This paper makes usage of the \citet{gilletal} modified Cholesky
factorizations for producing positive definite metric tensors from
the Hessian matrix for simplified manifold MALA methods. A new adaptive
step size procedure that resolves the shortcomings of metric tensors
derived from the Hessian in regions where the log-target has near
zero curvature in some direction is proposed. The adaptive step size
selection also appears to alleviate the need for different tuning
parameters in transient and stationary regimes. The combination of
the two constitute a large step towards developing reliable manifold
MCMC methods that can be implemented for models with unknown or intractable
Fisher information, and even for targets that does not admit a factorization
into prior and likelihood. Through examples of low to moderate dimension,
it is shown that proposed methodology performs very well relative
to alternative MCMC methods.

To handle high-dimensional problems, it is likely that a HMC variant
of the proposed methodology is needed. One avenue would be to make
$G(\mathbf{x})$ a smooth function via the usage of soft-max functions
\citep{1212.4693} in Algorithm \ref{alg:Left-looking-square-root}
and implement full Riemann manifold HMC along the lines of \citet{girolami_calderhead_11}.
This approach would enable the exploitation of sparsity of the Hessian
not afforded by methods based on spectral decompositions \citep{1212.4693},
but would still require computing third derivatives of $\log\tilde{\pi}$.
An interesting alternative that is currently being investigated involves
embedding HMC into a population MCMC framework where each member of
the population has the same target. In such a setup, one member of
the population is expended in each MCMC iteration for calculating
a position-specific mass matrix and time integration step size using
the modified Cholesky factorization and adaptive step size selection
procedure proposed here. These parameters are then applied in standard
HMC updates of the remaining population members to mimic the effects
of Riemann manifold HMC while retaining a computationally attractive
separable Hamiltonian. 

Another avenue for further work is to extend the adaptive step size
methodology via energy error arguments of Section \ref{sub:Line-search-mechanisms}
to other MCMC methods, via the observation that other, possibly non-symplectic,
numerical integration schemes applied to Hamilton's equations associated
with (\ref{eq:dummy_hamiltonian}) leads to different known proposal
distributions. In particular, a symplectic Euler type B integrator
\citep[page 26]{Leimkuhler:2004} lead to a $N(\mathbf{x}+\varepsilon^{2}G(\mathbf{x})^{-1}g(\mathbf{x}),\varepsilon^{2}G(\mathbf{x})^{-1})$
proposal which nests (for $\varepsilon=1$) the stochastic Newton
MCMC method of \citet{martin_etal2012}. A standard Euler integrator
lead to a position specific scale random walk proposal $N(\mathbf{x},\varepsilon^{2}G(\mathbf{x})^{-1})$.

\bibliographystyle{chicago}
\bibliography{/Users/Tore/owncloud/bibtex/kleppe}

\appendix

\section{A version of the \citet{gilletal} modified Cholesky\label{sec:A-version-of-chol}}

\begin{algorithm}
\begin{tabbing}
\hspace{2em} \= \hspace{2em} \= \hspace{2em} \= \\
{\bfseries Input}: $d\times d$ symmetric matrix $A$ and a small scale factor $u$.\\
step 0,1: $\tilde L \leftarrow I_d$.\\ 
step 0,2: $D_{j,j} \leftarrow A_{j,j},\;j=1,\dots d$.\\
step 0,3: $\nu \leftarrow \max_j |A_{j,j}|$.\\
step 0,4: $\xi\leftarrow \max_{i<j} |A_{i,j}|$.\\
step 0,5: $\phi^{2}\leftarrow \max(\nu,\xi/\sqrt{d^{2}-1},u)$.\\
step 0,6: $\delta \leftarrow u\max(\nu,\xi,1)$.\\
{\bfseries for} $j=1$ to $n$\\
\> step 1: {\bfseries if}$(j>1)$ $\tilde L_{j,k}\leftarrow \tilde L_{j,k}/D_{k,k},\;k=1,\dots,j-1$. \\
\> step 2: {\bfseries if}$(j<d)$ $\tilde L_{j+1:d,j} \leftarrow  A_{j+1:d,j}$. \\
\> step 3: {\bfseries if}$(1<j<d)$ $\tilde L_{j+1:d,j} \leftarrow \tilde L_{j+1:d,j} -(\tilde L_{j+1:d,1:j-1})(\tilde L_{j,1:j-1})^T$.\\
\> step 4: {\bfseries if}($j<d$) $\theta_j\leftarrow\max_{j<k\leq d} |\tilde L_{k,j}|$ {\bfseries else} $\theta_j\leftarrow 0$.\\
\> step 5: $D_{j,j} \leftarrow \max(\delta,|D_{j,j}|,\theta_j^2/\phi^2)$.\\
\> step 6: {\bfseries if}$(j<d)$ $D_{k,k} \leftarrow D_{k,k} - (\tilde L_{k,j})^2/D_{j,j}, k=j+1,\dots,d$.\\
{\bfseries end for}\\
{\bfseries Return} $\tilde L$ and $D$ (so that $\tilde L D \tilde L^T=A+J$) or $L=\tilde L D^{1/2}$ (so that $LL^T = A + J$).
\end{tabbing}

\protect\caption{\label{alg:Left-looking-square-root}A variant of the modified Cholesky
decomposition of \citet{gilletal}. }
\end{algorithm}
This section recaptures GMW's square root-free modified Cholesky algorithm
for finding lower triangular matrix $\tilde{L}$ with $\tilde{L}_{i,i}=1,\; i=1,\dots,d$
and diagonal matrix $D$ so that 
\[
\tilde{L}D\tilde{L}^{T}=A+J.
\]
The conventional lower triangular Cholesky factor given in (\ref{eq:Def_LLchol})
obtains as $L=\tilde{L}D^{1/2}$. GMW's approach takes as vantage
points:
\begin{enumerate}
\item A uniform lower bound on the diagonal elements of $D$ (or $L$),
i.e. $D_{j,j}=L_{j,j}^{2}\geq\delta,$ $j=1,\dots,d$ for a small
positive constant $\delta$. 
\item A uniform upper bound on the absolute off-diagonal elements of $L=\tilde{L}D^{1/2}$
that ensures that $\hat{A}$ is positive definite and numerically
stable, i.e. $|\tilde{L}_{i,j}\sqrt{D_{j,j}}|\leq\phi,\; i=2,\dots,d,\; j<i.$
\end{enumerate}
By taking $\phi^{2}=\max(\nu,\xi/\sqrt{d^{2}-1},u)$ where $\nu$
and $\xi$ are the maximal absolute diagonal- and off-diagonal elements
respectively, \citet{gilletal} show that $\hat{A}$ is positive definite
while at the same time a bound on the infinity norm of $J$ is minimized.
Throughout this paper, $\delta$ is taken to be $u\max(\nu,\xi,1)$.
The scale factor $u$ is commonly taken to be close to machine accuracy
for optimization applications, but in this work it is left as tuning
parameter.

GMW's approach for implementing the above bounds is based on a square
root free Cholesky factorization \citep[p. 146]{noce:wrig:1999} and
is given in Algorithm \ref{alg:Left-looking-square-root}. \citet{gilletal}
observed that immediately after step 3 in the $j$th iteration of
the algorithm, the $j$th column of the unfinished $\tilde{L}$ contains
$D_{j,j}$ times the $j$th column of the final $\tilde{L}$. Therefore
$D_{j,j}$ can be chosen (step 4 and 5) during the course of the algorithm
to respect the two bounds given above. The difference in $D_{j,j}$
between after and before step 4 in iteration $j$ amounts to the $j$th
diagonal element of $J$, and simply removing steps 4 and 5 corresponds
to a standard standard square root free Cholesky algorithm. The modifications
essentially adds $O(d^{2})$ comparisons to the computational complexity
relative to the standard dense Cholesky algorithm it derives from,
and is therefore asymptotically insignificant comparing to the $O(d^{3}/6)$
floating point operations needed to carry out the latter.

\section{Proof of Proposition 1\label{sec:Proof-of-Proposition}}

Irreducibility of the update of $\mathbf{x}$ in steps 1-4 is ensured
by the fact that the proposal distribution is Gaussian with finite
and positive definite covariance matrix whenever $0<\varepsilon(\mathbf{x},\mathbf{w})<\infty$
for all $(\mathbf{x},\mathbf{w})$ in the support of $\pi(\mathbf{x},\mathbf{w})$.
The update of $\mathbf{w}$ in step 5 is trivially irreducible since
it amounts to iid sampling. Consequently the overall update in steps
1-5 is irreducible as any point on the support of the target is attainable
in each iteration. 

Aperiodicity of the update of $\mathbf{x}$ is ensured by the fact
that it involves a non-trivial accept-reject in step 4 \citep[section 7.3.2]{robert_casella}
and aperiodicity of the update of $\mathbf{w}$ is trivial since it
it iid. Consequently, the overall update in steps 1-5 is thus aperiodic.

To verify that $\pi(\mathbf{x},\mathbf{w})$ is the invariant distribution
of steps 1-5, first observe that for each given $\mathbf{w}_{t}$,
steps 1-4 define a reversible MH step with $\pi(\mathbf{x})$ as invariant
distribution. The reversibility is consequence of the fact that $\varepsilon_{f},\varepsilon_{b}$
are computed with the same $\mathbf{w}$-argument, and therefore within
steps 1-4 of each iteration $\varepsilon(\mathbf{x},\mathbf{w})$
is effectively a function of the first argument only. Denote by $A(\mathbf{x}_{t+1}|\mathbf{x}_{t},\mathbf{w}_{t})$
the transition kernel associated with steps 1-4. Then the overall
transition kernel of steps 1-5 may be written as $B(\mathbf{x}_{t+1},\mathbf{w}_{t+1}|\mathbf{x}_{t},\mathbf{w}_{t})=A(\mathbf{x}_{t+1}|\mathbf{x}_{t},\mathbf{w}_{t})\pi(\mathbf{w}_{t+1})$.
That $\pi(\mathbf{x},\mathbf{w})$ is the invariant distribution of
$B$ is then seen via
\begin{eqnarray*}
\int\int\pi(\mathbf{x},\mathbf{w})B(\mathbf{x}^{\prime},\mathbf{w}^{\prime}|\mathbf{x},\mathbf{w})d\mathbf{x}d\mathbf{w} & = & \pi(\mathbf{w}^{\prime})\int\pi(\mathbf{w})\int A(\mathbf{x}^{\prime}|\mathbf{x},\mathbf{w})\pi(\mathbf{x})d\mathbf{x}d\mathbf{w},\\
 & = & \pi(\mathbf{w}^{\prime})\pi(\mathbf{x}^{\prime})=\pi(\mathbf{x}^{\prime},\mathbf{w}^{\prime}),
\end{eqnarray*}
where the second equality follows from the reversibility of steps
1-4 for each $\mathbf{w}$. 
\end{document}